\documentclass[a4paper,11pt]{article}
\usepackage{jheppub} 
\usepackage{graphicx}
\usepackage{dcolumn}
\usepackage{bm}
\usepackage[usenames,dvipsnames]{color}
\usepackage[svgnames]{xcolor}
\usepackage{soul}
\usepackage{subcaption}
\usepackage{enumitem}
\usepackage{xspace}
\usepackage{booktabs}
\usepackage{siunitx} 
\usepackage[ISO]{diffcoeff}
\usepackage{subcaption}
\usepackage{ragged2e}
\usepackage{mathrsfs}
\usepackage{hyperref}
\usepackage{cleveref}
\usepackage[normalem]{ulem}
\usepackage{amsmath}
\usepackage{wrapfig}
\usepackage{esvect} 
\usepackage{titlesec}
\usepackage{titletoc}
\usepackage{chngcntr}
\usepackage{fontawesome5}
\usepackage{mleftright}
\usepackage[nodayofweek]{datetime}
\usepackage{tikz}
\usepackage{tabularx}

\DeclareSIUnit\year{yr} 
\def\arraystretch{2}\tabcolsep=10pt

\DeclareCaptionJustification{justified}{\justifying}
\definecolor{firebrick}{HTML}{B22222}

\hypersetup{
    colorlinks=true,       
    linkcolor=firebrick,          
    citecolor=firebrick,        
    filecolor=firebrick,      
    urlcolor=firebrick           
}

\newcommand{\dd}{\mathop{}\!\mathrm{d}}
\newcommand{\ee}{\mathrm{e}}
\newcommand{\mrm}[1]{\mathrm{#1}}
\newcommand{\usim}{\mathord{\sim}}
\newcommand{\boldvec}[1]{\boldsymbol{#1}}
\newcommand{\rvec}{{\bm r}}
\newcommand{\rhat}{\hat{\rvec}}

\newcommand{\Fvec}{{\bm F}}
\newcommand{\dvec}{{\bm \nabla}}
\newcommand{\nvec}{{\bm n}}
\newcommand{\nhat}{\hat{\nvec}}
\newcommand{\xhat}{\hat{\bm x}}
\newcommand{\zhat}{\hat{\bm z}}
\newcommand{\Windchime}[1]{\text{\sc Windchime}}

\newcommand{\MPl}{M_\mrm{Pl}}
\newcommand{\MP}{\MPl}

\newcommand{\rref}[1]{ref.~\cite{#1}}
\newcommand{\Rref}[1]{Ref.~\cite{#1}}

\titleclass{\mysection}{straight}[\section]
\titleformat{\mysection}[runin]
  {\itshape}{\thesection}{}{}[.---]
\titlespacing{\mysection}{1em}{1em}{0em}

\title{\boldmath Towards the Direct Detection of Composite Ultraheavy Dark Matter in Quantum Sensor Arrays}
\author[a,b]{Dorian W.~P.~Amaral,}
\emailAdd{damaral@ifae.es}
\author[a,c]{\ Erqian Cai,} 
\emailAdd{erqian.cai@stonybrook.edu}
\author[a]{\ Andrew~J.~Long,}
\emailAdd{andrewjlong@rice.edu}
\author[a]{\ Juehang Qin,}
\emailAdd{jq8@rice.edu}
\author[a]{\ and \ Christopher D. Tunnell}
\emailAdd{tunnell@rice.edu}

\affiliation[a]{Department of Physics and Astronomy, Rice University, \\ 6100 Main Street, Houston, TX 77005, U.S.A.}
\affiliation[b]{Institut de F\`{ı}sica d’Altes Energies (IFAE), The Barcelona Institute of Science and Technology,
Campus UAB, 08193 Bellaterra (Barcelona), Spain}
\affiliation[c]{Department of Physics and Astronomy at Stony Brook University, \\ 100 Nicolls Road, Stony Brook, NY 11794, USA}

\abstract{Quantum sensor arrays have recently been proposed as a promising platform for the direct detection of ultraheavy dark matter, which is typically assumed to behave as a point-like particle. However, particles with masses at or above the Planck scale cannot be elementary; instead, they must exist as composite objects with finite spatial extent. Such spatially extended dark matter models lead to distinctive phenomenology in these detectors, particularly when the dark matter also interacts through long-range forces with their own characteristic length scales. In this work, we study the sensitivity of quantum sensor arrays to composite, ultraheavy dark matter interacting via both gravity and a novel Yukawa force. We consider three phenomenologically motivated density profiles---a tophat, a Gaussian, and an exponential---and contrast their signals with the point-like limit. Using a Monte Carlo analysis based on the predicted impulse signals and estimates of thermal and quantum noise, we obtain sensitivity projections for a future realization of a quantum sensor array.  We find a non-trivial interplay between the dark-matter scale radius, the inter-sensor spacing, and the Yukawa screening length.  Future accelerometer arrays would provide valuable information about the mass and size of composite ultraheavy dark matter, and our work will help to characterize the signatures of different theoretical models of ultraheavy dark matter. 
}


\begin{document}
\maketitle
\flushbottom

\section{Introduction}
\label{sec:intro}

Dark matter (DM) is a mysterious, pervasive substance that has thus far revealed itself only through its gravitational interactions. Typically, the cold DM is assumed to be an elementary particle, informing terrestrial DM searches. However, dark matter could instead exist as a composite, and therefore spatially extended, \textit{clump} that is formed from more elementary constituents.
Compellingly, if these extended particles possess net masses at or above the Planck scale, $\MP \equiv (\hbar c / G)^{1/2} \approx 1.2 \times 10^{19}\,\mrm{GeV}/c^2 \approx 22\,\mrm{\mu g}$, then they must be composite if they are not to be black holes. Dark matter candidates with such high masses are often referred to as ultraheavy dark matter (UHDM), and in this work we are interested in the effect of their finite spatial extent on the prospects for direct detection using a future array.  

Ultraheavy dark matter is expected to arise in many theoretical frameworks~\cite{Carney:2022gse}.  
A few notable examples include primordial black holes~\cite{Carr:2021bzv}, quark nuggets/stranglets~\cite{Witten:1984rs,Farhi:1984qu}, six-flavor quark nuggets~\cite{Bai:2018vik}, dark quark nuggets~\cite{Bai:2018dxf}, axion quark nuggets~\cite{Zhitnitsky:2021iwg}, Fermi-balls~\cite{Hong:2020est}, $Q$-balls~\cite{Kusenko:2001vu}, dark blobs~\cite{Grabowska:2018lnd}, MACROs~\cite{Jacobs:2014yca}, Planck star remnants~\cite{Trivedi:2025vry}, asymmetric dark matter nuggets~\cite{Wise:2014ola,Wise:2014jva,Gresham:2018anj}, and WIMPzillas~\cite{Kolb:2023ydq}. The landscape of models is vast and varied; the dark matter's mass, for instance, could be below that of a proton or above that of a star. We restrict our attention to composite UHDM with masses $M \sim \MP$.  

The motivation for this narrow focus is twofold.  
Firstly, we are interested in the feasibility of ultraheavy DM direct detection for laboratories on Earth as this has historically been viewed by experimentalists to be challenging, if not impossible.  
This imposes an upper limit of around the Planck mass if we are to measure an appreciable flux of DM particles within a $\usim 1\,\mrm{m}$ scale terrestrial experiment. Secondly, the Standard Model of  elementary particles is an effective field theory that is expected to break down at (or before) the Planck scale, where new particles and forces should enter to complete the theory.  
Therefore, it is natural to think of these Planck-scale particles as stable dark matter candidates, making $M \approx \MP$ an especially theoretically-favored mass regime.  

In this work, we call attention to the fact that we should expect UHDM with masses $M\gtrsim\MP$ to be composite and spatially extended.  
The Planck scale, defined by combining Newton's gravitational constant $G$ with Planck's constant $\hbar \equiv h/2\pi$ and the speed of light $c$, stands for the merger of gravity, quantum mechanics, and relativity.  
It is common to see the Planck mass arising in the study of black holes, as it relates their mass $M_\mrm{BH}$ to their Schwarzschild radius $R_\mrm{BH}$ through $R_\mrm{BH} = 2 \hbar M_\mrm{BH} / \MP^2 c$.  
Since an object with mass $M > \sqrt{\pi} \MP$ would have a smaller Compton wavelength $\lambda = 2 \pi \hbar / M c$ than the Schwarzschild radius of an equal mass black hole, it is reasonable to postulate that such an object must be either a black hole or a non-elementary composite object possessing a finite spatial extent.

However, experimental searches for UHDM in laboratories on Earth often assume that dark matter is point-like.  
For instance, xenon-based direct detection experiments such as XENON1T~\cite{XENON:2023iku} and LZ~\cite{LZ:2024psa} have constrained point-like UHDM up to masses $M \sim 10^{17}~\mrm{GeV}$ in this way (see also \cite{Digman:2019wdm}). A variety of novel technologies have been proposed to search for UHDM. Examples include three-dimensional mechanical sensor arrays~\cite{Carney:2020xol}, hydrophones~\cite{Cleaver:2025etu}, gradiometers~\cite{Badurina:2025xwl}, and cloud chambers~\cite{Gao:2025ryi}. In this work, we focus on quantum sensor arrays, wherein an array can be conceivably constructed using a variety of sensor technologies, including but not limited to optically levitated sensors, magnetically levitated sensors, and torsion balances.  We characterize these as ``quantum sensor arrays'' since reaching the desired sensitivity for dark matter detection would require pushing beyond the standard quantum limit.
This experimental platform is rapidly developing in both scale and sensitivity, and it is a good choice to search for Planck-scale dark matter via long-range interactions. This is because it is feasible to construct such arrays at the $\usim 1\,\mrm{m}$ scale, allowing us to accumulate the necessary statistics for a detection despite the low expected rate of DM at these high masses ($\usim 0.2\,\mrm{yr}^{-1}$).

Prior work by some of the authors studied the prospects for detecting UHDM with a three-dimensional array of accelerometers under the assumption that the dark matter size is negligible compared to the $\usim\mrm{cm}$ spacing between the sensors~\cite{Qin:2025jun}.  
However, the conclusion that UHDM above the Planck mass can have a spatial extent is inescapable, and we therefore revisit the phenomenology underpinning this detector concept in this paper. We begin by stating our model assumptions for UHDM in \cref{sec:modeling}, where we introduce the concept of a DM \textit{clump}, as well as the density profiles and Yukawa parametrization we consider in this work. We then move on to describing the detector paradigm of the quantum sensor array in \cref{sec:experiment}, outlining its basic working principles, computing the signal-to-noise expected at such a detector from a passing UHDM clump, and describing the Monte Carlo numerical simulation procedure we employ in drawing our sensitivity projections. Finally, we present our results in \cref{sec:results}, showing how the sensitivity is optimized when certain length scales match and how it degrades as the dark matter clump scale grows beyond that of the sensor array. Our results can be used to advise future detector design in searching for ultraheavy dark matter.

\section{Modeling Composite Ultraheavy Dark Matter}
\label{sec:modeling}

To predict the signals expected at a sensor array, we must specify both the spatial distribution of mass within a composite dark matter clump and the nature of its interactions with ordinary matter. In this section, we introduce the phenomenological framework that captures these properties.

\subsection{Ultraheavy Dark Matter Clumps}
\label{subsec:properties}

An ultraheavy dark matter \textit{clump} can be completely characterized by three properties: its mass $M$, a characteristic length scale $R$, and a mass density profile $\rho(\boldvec{r})$ connecting these quantities. We assume that the DM profile is radially symmetric, such that $\rho(\boldvec{r}) = \rho(r)$ with $r \equiv |\boldvec{r}|$.  The differential mass $\dd M$ contained within a differential radius $\dd r$ is then related to the density profile  via $\dd M  = 4 \pi r^2 \rho(r)\dd r$. Combined with an assumed interaction potential, these three quantities are sufficient for us to predict the signals that we would expect to see from these extended DM candidates and allow us to assess the sensitivities of accelerometer arrays.  

\begin{figure}[t]
    \centering
    \includegraphics{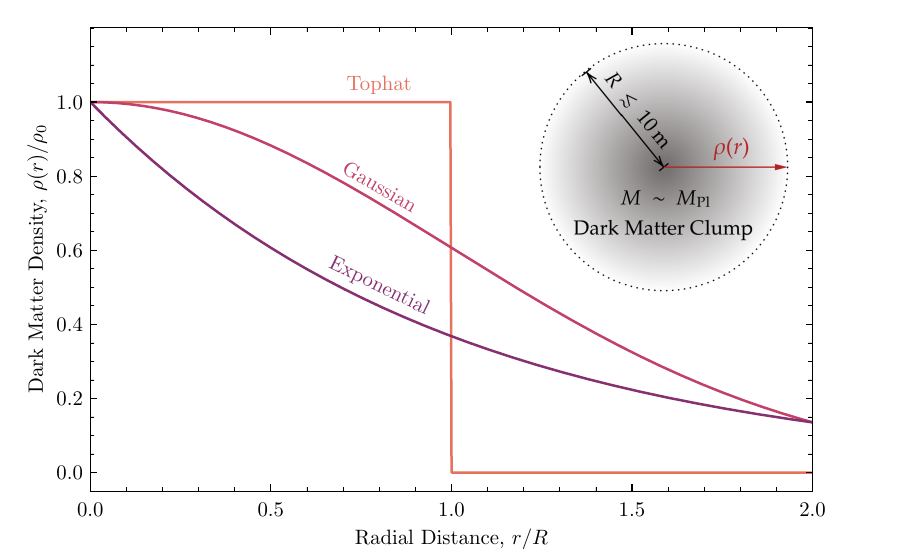}
    \caption{\label{fig:profiles}
    The normalized mass density profiles of the dark matter clump $\rho(r)/\rho_0$ with radial distance $r$ as a fraction of the clump's characteristic radius $R$. The considered spatially extended profiles are the tophat, Gaussian, and exponential profiles, as given in \cref{eq:rho_profs}. Schematically shown is also a dark matter clump, considered to have a mass $M \sim \MP$ and radius $R \lesssim 10\,\mrm{m}$.
    }
\end{figure}

Different UHDM candidates can have different mass density profiles~\cite{Bai:2020jfm}. For instance, dark quark nuggets, and more generally Fermi-balls, have an approximately constant energy density in their interiors---this results in a `tophat' mass density profile~\cite{Witten:1984rs, Zhitnitsky:2002qa,Bai:2018dxf,Bai:2018vik,Bai:2020jfm,Hong:2020est}. On the other hand, dilute axion stars are stabilized via the equilibrium of the inward gravitational and outward quantum pressures (similar to white dwarfs and neutron stars), leading to an exponential density profile~\cite{Kolb:1993zz,Bai:2020jfm}. A Gaussian profile can further act as a phenomenological intermediate between these. Together with the typically assumed point-like profile, we thus explore four functional forms for $\rho(r)$:
\begin{subequations}\label{eq:rho_profs}
\begin{alignat}{2}
    \rho_\mrm{P}(r) & \equiv M \delta^{3}(r) 
    \qquad&&\text{(Point-like)}\,,\\[2ex] 
    \rho_\mrm{T}(r) & \equiv \begin{cases}
    \frac{3 M}{4\pi R^3} \quad &\text{if} \quad r \leq R \\ 
    0 \quad &\text{else}
    \end{cases} 
    \qquad&&\text{(Tophat)} \,, \\[2ex] 
    \rho_\mrm{G}(r) & \equiv \frac{M}{(2 \pi R^2)^{3/2}} \, \ee^{-r^2 / 2 R^2} 
    \qquad&&\text{(Gaussian)} \,, \\[2ex]  
    \rho_\mrm{E}(r) & \equiv \frac{M}{8 \pi R^3} \, \ee^{-r/R} 
    \qquad&&\text{(Exponential)} \,.
\end{alignat}
\end{subequations}
Each of the normalizations above ensure that we retrieve the total mass of the particle upon volume integration. We visualize these profiles in \cref{fig:profiles}, along with a schematic of a dark matter clump.

\subsection{Dark Matter Interactions}
\label{subsec:interactions}

Having established the density profile of a dark matter clump, we now specify how it interacts with the sensors in our proposed detector. \Rref{Qin:2025jun} showed that state-of-the-art mechanical sensor arrays are insensitive to (point-like) DM interacting via gravity alone. Instead, an additional interaction is required, and a Yukawa force is a natural choice for this~\cite{Blanco:2021yiy,Grabowska:2018lnd,Jacobs:2014yca}. Relatedly, \rref{Xu:2025xaq} also considered Yukawa interactions for spatially extended DM, exploring how imbuing DM at the WIMP mass regime (around $\usim\mrm{GeV}$) with a finite extent can impact direct detection constraints.
In this work, we assume that a DM clump, having a mass density profile $\rho$, couples via this fifth force proportionally to the mass of a test object. The total potential energy of a test mass $m_s$ at a time-varying position $\boldvec{r}(t)$ due to the clump can then be written as~\cite{Fischbach:1992fa,Adelberger:2003zx} 
\begin{align}\label{eq:potential_energy}
    U\textbf{(}\rvec(t)\textbf{)} = - G m_s \int \! \dd^3 \rvec^\prime \, \frac{ \rho(r^\prime)}{|\rvec(t) - \rvec^\prime|} \Bigl( 1 + \alpha \, \ee^{- |\rvec(t) - \rvec^\prime| / \lambda} \Bigr) 
    \;.
\end{align}
Here, $G$ is Newton's universal gravitational constant, $\alpha$ is a dimensionless parameter governing the strength of the fifth force relative to that of gravity, and $\lambda$ is the screening length of the force. The volume integral is taken over all space with respect to the rest frame coordinates of the DM clump.

Throughout our sensitivity analysis, we will work in the rest frame of the dark matter clump such that a fixed test particle in the laboratory frame moves relative to it. This gives the position vector of the test mass a time dependence via $\boldvec{r}(t) \equiv \boldvec{r}_0 - \boldvec{v}_\mrm{DM} t$, where $\boldvec{r}_0$ is some initial position vector for the test mass relative to the center of mass of the clump and $\boldvec{v}_\mrm{DM}$ is its associated velocity vector in this frame. Since the DM clump moves non-relativistically, $|\boldvec{v}_\mrm{DM}| \ll c$, we neglect relativistic effects.

We can perform the angular part of the integral in \cref{eq:potential_energy} by exploiting the angular symmetry of the integrand. The total force on a point-like test mass due to both gravity and a novel Yukawa force may then be written as
\begin{align}\label{eq:Fvec}
    \Fvec\textbf{(}\rvec(t)\textbf{)}  & = - \dvec U\textbf{(}\rvec(t)\textbf{)}  = - \frac{4\pi G m_s}{r} \hat{\boldvec{r}} \int_0^{\infty} \! \dd r' \rho(r') \, r' \, \mathcal{I}(r, r',t)\,,
\end{align}
where we have defined the function $\mathcal{I}(r, r',t)$ as
\begin{equation}
    \mathcal{I}(r, r', t) \equiv
    \begin{cases}
    \frac{r^\prime}{r}\left[\frac{r^\prime}{r} - \frac{\alpha}{2} \left(1 + \frac{\lambda}{r}\right)\left(\ee^{-(r + r')/\lambda} - \ee^{-(r - r')/\lambda}\right)
    \right] \quad &\text{if $r^\prime \leq r$}\,, \\[2ex] 
    - \frac{\alpha}{2}\left[\left(1 + \frac{\lambda}{r}\right)\ee^{-(r + r')/\lambda} + \left(1 - \frac{\lambda}{r}\right)\ee^{(r - r')/\lambda}\right] \quad &\text{if $r^\prime > r$} \,.
    \end{cases}
\end{equation}
One feature present in the Yukawa force and absent from the gravitational force is that mass beyond a given radial position $r$ can contribute a net effect to it. The presence of the exponential suppression term leads to a non-inverse-square-law behavior, spoiling the result of Newton's shell theorem. Nevertheless, this result is retrieved in the limit that $\lambda \rightarrow \infty$, corresponding to a force, like gravity, with an infinitely large scale.

The form for the force given in \cref{eq:Fvec} is general, and it can be used for any well-behaved and  spherically symmetric density profile. However, for the density profiles we consider in \cref{eq:rho_profs}, the remaining radial integral can be evaluated analytically, albeit in terms of special functions. We supply these results in \cref{sec:app_forces} for completeness.

\section{Direct Detection with a Quantum Sensor Array}
\label{sec:experiment}


The rapid pace of technology development for quantum sensing has opened up a new measurement frontier. Experimental setups based on accelerometers have recently been proposed as promising direct probes of DM~\cite{Carney:2020xol,Manley:2020mjq,Monteiro:2020wcb,Afek:2021vjy,Antypas:2022asj,Brady:2022qne,Buchmueller:2022djy,Windchime:2022whs,Higgins:2023gwq,Chou:2023hcc,Beckey:2023shi,Kalia:2024eml,Amaral:2024tjg,Amaral:2024rbj,Qin:2025jun}. \cite{Carney:2019pza} was the first to propose an array built from such sensors to search for UHDM, suggesting that it would offer sufficient sensitivity to detect Planck-scale dark matter via the gravitational interaction alone---from this was borne the \Windchime{} project~\cite{Windchime:2022whs}. A follow-up work, however, showed that the required level of quantum noise reduction appears prohibitive for a purely gravitational detection, requiring upwards of $85\,\mrm{dB}$ of quantum noise suppression before reaching a thermal noise floor permitting such a detection~\cite{Qin:2025jun}. Nevertheless, as we alluded to above, if UHDM is allowed to interact via an additional fifth force, 3D sensor arrays can indeed have leading sensitivities, motivating the parametrization we use in our analysis~\cite{Monteiro:2020wcb,Blanco:2021yiy,Qin:2025jun}. Devices based on magnetic levitation technology have also been proposed as a promising way to detect the existence of such fifth forces mediating interactions between standard model matter~\cite{Amaral:2025zgk}.  

\subsection{Sensor Array}
\label{subsec:array}

\begin{figure}[t]
    \centering
    \input{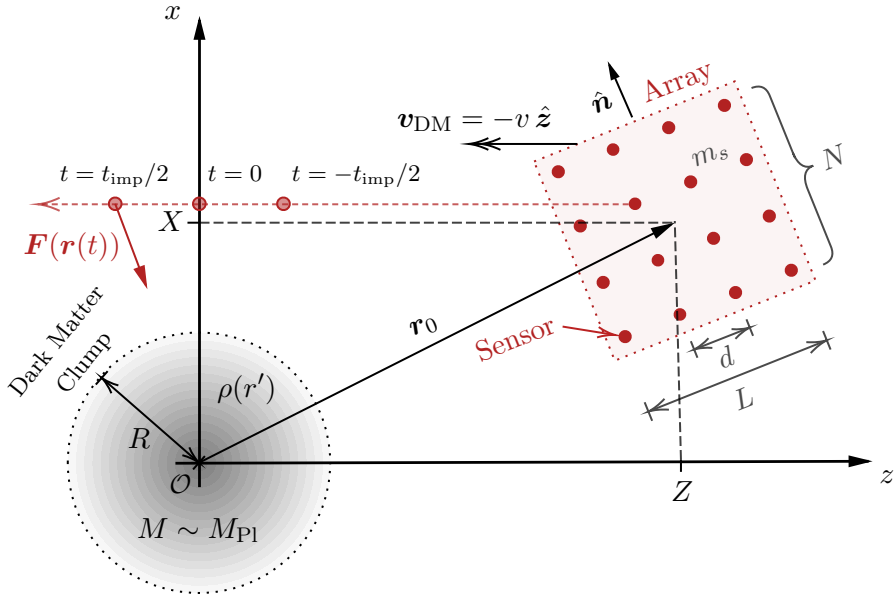}
    \caption{\label{fig:array}
    A cross section of the $3$D quantum sensor array in the rest frame of the dark matter clump. The clump is placed at the origin $\mathcal{O}$ of the $xz$--coordinate plane and has a mass $M$. Its mass density profile is governed by the function $\rho(r')$, which also depends on an assumed characteristic length scale $R$ for the DM clump. The sensor array is placed at an initial central displacement $\boldvec{r}_0 \equiv X\,\hat{\boldvec{x}} + Z\, \hat{\boldvec{z}}$, where $X$ and $Z$ are random variables, with a random orientation $\hat{\boldvec{n}}$. In this frame, the array moves at a velocity $\boldvec{v}_\mrm{DM} = - v\,\hat{\boldvec{z}}$, with $v$ a random variable distributed according to a 3D Maxwell-Boltzmann distribution with a scale parameter $v_0 / \sqrt{2}$ and $v_0 \approx 238\,\mrm{km\,s^{-1}}$ the dark matter virial velocity. The sensor array consists of a total of $N \times N \times N$ sensors of mass $m_s$, where $N$ is the number of sensors along any one dimension. The array has a total length $L = (N-1)d \sim 1\,\mrm{m}$, where $d \sim 10\,\mrm{cm}$ is the inter-sensor spacing. Each sensor experiences a force $\boldvec{F}\textbf{(}\boldvec{r}(t)\textbf{)}$ due to the dark matter clump. Also shown are the positions of a single sensor at times $t = -t_\mrm{imp}/2$, $t = 0$, and $t = t_\mrm{imp}/2$, where $t_\mrm{imp}$ is the typical duration of an impulse.
    }
\end{figure}

We imagine a cubic array built from a total of $N \times N \times N$ sensors of mass $m_s$, which are separated by a distance $d$ in any given axis. Since the number of sensors along a particular direction is $N$, the length of the array is $L = (N-1
) d$. In the rest frame of the DM clump, the center of the sensor array is placed at an initial displacement $\boldvec{r}_0 \equiv X \hat{\boldvec{x}} + Z \hat{\boldvec{z}}$, where $X$ and $Z$ are random coordinate variables. The symmetry of the DM clump allows us to rotate the coordinate axes such that we only need to consider the motion of the array within a plane, and we arrange these axes such that the array moves at a velocity $\boldvec{v}_\mrm{DM} = - v \hat{\boldvec{z}}$, with $v$ also a random variable. We elaborate further on these parameters in \cref{subsec:sim}, where we describe our Monte Carlo procedure to draw our projected sensitivities. We visualize this scenario in \cref{fig:array}, showing a cross section of our proposed array setup in the presence of a dark matter clump.

Each sensor in the array is assumed to be identical and to possess a single sensitivity axis pointing in a common, randomized direction $\hat{\boldvec{n}}$. Along this sensitivity axis, quantities such as displacement and acceleration can be measured. Examples of such axial sensors include optomechanical cavities~\cite{Aspelmeyer:2013lha} and micro-electro-mechanical systems (MEMS)~\cite{Krause:2012iud,Errando-Herranz:2022sik}; nevertheless, other systems with access to more degrees of freedom similar when considering additional modes. Going beyond our study, one advantage of using multiple degrees of freedom is the increased sensitivity to the directionality of a dark matter `track' through the sensor; however, since we ultimately will only consider a summed signal-to-noise ratio argument, we do not take advantage of this here.

Given a particular array realization, the rate at which dark matter clumps cross it is an important quantity for a sensitivity estimate since it dictates the feasibility of our detection scheme. For an array of length $L$ and cross-sectional area $A \sim L^2$, the expected DM crossing rate is
\begin{equation}\label{eq:rate}
    \mathcal{R} = n_\mrm{DM} v_0 A \simeq 0.2\,\mrm{yr^{-1}} \left(\frac{\MP}{M}\right)\left(\frac{L}{1\,\mrm{m}}\right)^2\,,
\end{equation}
where $n_\mrm{DM} \equiv \rho_\mrm{DM} / M$ is the local dark matter number density, and $\rho_\mrm{DM} \approx 0.3\,\mrm{GeV\,cm^{-3}}$ the local dark matter mass density. Since a reasonable exposure is a few years, the Planck mass sets a natural upper limit on the range of masses to which our array can be sensitive. 

As a dark matter clump passes through the array, or equivalently as the array moves through the dark matter clump in the DM center-of-mass frame, the clump imparts a force $\boldvec{F}\textbf{(}\boldvec{r}(t)\textbf{)}$ on a particular sensor. This leads to a series of impulses throughout the array, causing the sensors to oscillate at their natural frequency $\omega_0$. Due to the coupling between the mechanical modes and the environment, these oscillations will be dampened, resulting in a gradual transfer of energy from the system at the mechanical damping rate $\gamma$.   The mechanical quality factor $Q \equiv \omega_0 / \gamma$ captures the decay of these oscillations, and it is approximately equal to the number of oscillation cycles per $e$-folding of the energy loss. Mechanical accelerometers can be of very high quality, with quality factors in excess of $Q \gtrsim 10^6$ and even $Q \gtrsim 10^{10}$, as can be the cases for magnetically and electrically levitated systems, respectively~\cite{Fuchs:2023ajk, 
Dania:2023uyz}.

In modeling the sensor response, we treat the sensors as point-like. 
This is a good approximation as long as the separation between the center of the dark matter clump and the center of the sensor is larger than both the clump radius and the sensor radius $R_s$.  
Thus, this approximation fails if the distance of closest approach $b$ is below the sensor radius, at which point the force imparted on the sensor can increase to very large values. To remedy this, we approximately treat the finite size of a sensor by capping the force such that $\Fvec(\rvec)$ is evaluated at $|\rvec| = R_s$ if $|\rvec| < R_s$. For reference, a $100\,\mrm{g}$ sensor treated as a homogeneous ball of high-density material, such as lead $\rho_\mrm{Pb} \approx 11.34\,\mrm{g\,cm^{-3}}$, has a radius of $R_s \approx 1 \, \mathrm{cm}$.

The amount of time $t_\mrm{imp}$ that elapses while the DM clump delivers an impulse depends on the clump's characteristic size $R$, the Yukawa screening length $\lambda$, the speed of the clump $v$, and the impact parameter $b$. The last two of these quantities are random, with the former depending on the assumed DM halo model and the latter on the path the clump takes as it traverses the array. The time $\tau$ over which a measurement is made encapsulates both the duration of a measurement and the time interval between subsequent measurements. In \cref{subsec:signal}, we will discuss how there is an optimal choice for the measurement time $\tau_\mrm{opt}$ that minimizes the amount of noise introduced into the measurement. Nonetheless, for the parameters of interest in this work, it is generally true that $t_\mrm{imp} \ll \tau_\mrm{opt} \ll 2\pi / \omega_0$. Thus, the measurement time is such that we can completely capture an impulse signal, and we are able to fully observe the oscillatory response of each sensor. We are therefore free to choose $\tau = \tau_\mrm{opt}$ \cite{Qin:2025jun}.  

\subsection{Impulse Sensing with a 3D Array}
\label{subsec:array}

\subsubsection{Impulse Signal of a Single Sensor}
\label{subsec:signal}

As a DM clump moves through the sensor array, it imparts a series of impulses on the component sensors. We take the signal $\mathcal{S}_i$ on the $i^\mrm{th}$ sensor, with $i \in \{1, 2,\dots, N^3\}$, to be the absolute value of the impulse delivered to it by the passing DM clump. We take the absolute value to ensure that contributions from sensors on opposite sides of an array do not cancel. In the rest frame of the clump and for a sensor with sensitivity axis pointing along $\hat{\boldvec{n}}$, this is given by 
\begin{equation}\label{eq:sig}
    \mathcal{S}_i \equiv \left\lvert\int_{-\infty}^{\infty} \! \dd t \, \Fvec\bm{(}\rvec_i(t)\bm{)} \cdot \nhat \right\lvert\,,
\end{equation}
where $\Fvec\bm{(}\rvec(t)\bm{)}$ is given in \cref{eq:Fvec} and $\rvec_i(t)$ is the position vector of the $i^\mrm{th}$ sensor at time $t$. This position may be written as $\rvec_i(t) = \rvec_{0,i} - v t\hat{\boldvec{z}}$, where $\rvec_{0,i}$ is its initial position and $v$ is its speed in the clump frame. In practice, we compute this integral numerically by considering time steps much smaller than the typical impulse duration $t_\mrm{imp}$ and initiating the array far away from the clump; this allows us to fully capture the expected signal. 

\begin{figure}[t!]
    \centering
    \includegraphics{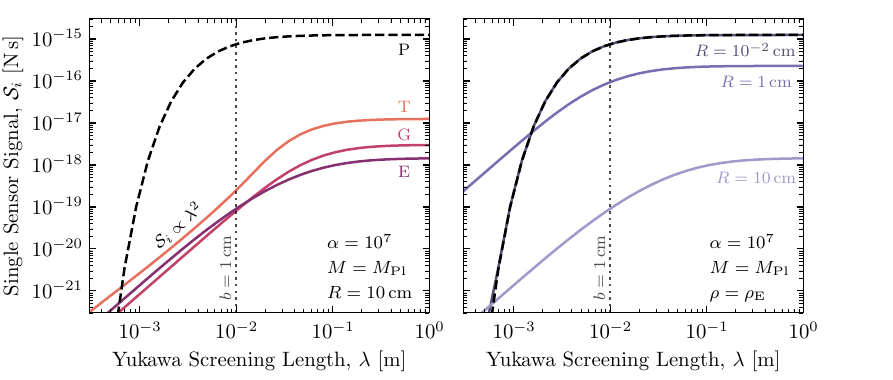}
    \caption{The single sensor signal $\mathcal{S}_i$, defined as the impulse delivered to a sensor by a passing DM clump, with varying Yukawa screening length $\lambda$. The signal is calculated via \cref{eq:sig}, taking the Yukawa interaction strength to be $\alpha = 10^{7}$ and assuming that a sensor of mass $m_s = 100\,\mrm{g}$ passes by a DM clump of mass $M = \MP$ with impact parameter $b = 1\,\mrm{cm}$. \textbf{Left:} The signal for the point-like ($\mathrm{P}$), tophat ($\mathrm{T}$), Gaussian ($\mathrm{G}$), and exponential ($\mathrm{E}$) dark matter mass profiles (cf.~\cref{eq:rho_profs}) for a clump radius of $R = 10\,\mrm{cm}$. For $\lambda \lesssim b$, the point-like signal decays exponentially, while that for any one of the extended profiles scales as $\mathcal{S}_i \propto \lambda^2$. \textbf{Right:} The signal for the exponential profile with varying clump radius. As $R \rightarrow 0$, we retrieve the point-like result.}
    \label{fig:single_sig}
\end{figure}

In \cref{fig:single_sig}, we show the predicted single sensor signal $\mathcal{S}_i$ with varying Yukawa screening length $\lambda$ and a fixed interaction strength of $\alpha = 10^{7}$. We take the DM clump to have a mass of $M = \MP$, and we consider a sensor of mass $m_s = 100\,\mrm{g}$. The right panel of \cref{fig:single_sig} shows how by decreasing this radius towards zero, we can retrieve the point-like result. In the left panel of \cref{fig:single_sig}, we show the signals for each of our profiles for a clump radius of $R = 10\,\mrm{cm} < b$, such that the sensor passes through the clump. This result highlights a recurring observation we will see in this work: the point-like signal is stronger than the signals predicted from finite-sized DM clumps for large values of $\lambda$, whereas it is weaker than the clump signals for small values of $\lambda$.

When $\lambda \gg R$, the fifth force is effectively unsuppressed over the scale at which the impulse is mostly delivered, $r \sim b$. The effect is that the force acts similarly to gravity, with only the total enclosed mass within radial distance $r$ contributing over the sensor trajectory. The signal then becomes independent of $\lambda$, with the transverse force approximately given by $F_\perp \sim G M(<\!r) m_s b / r^3$, where $M(<\!r)$ is the mass within radius $r$. Taking $M(<\!r) \sim \rho_0 r^3$, with $\rho_0$ the normalization of the profile, and appreciating that the Yukawa force will act over a time $\Delta t \sim 2R / v$, this yields a signal $\mathcal{S}_i \sim F_\perp \Delta t \sim G \rho_0 m_s  b R / v$. 

This relation helps to explain two observations. Firstly, taking $\rho_0 \sim M / R^3$, we see that $\mathcal{S}_i \propto R^{-2}$, indicating that larger clump radii tend to decrease the signal, as seen in the right panel of \cref{fig:single_sig}. Secondly, since $\mathcal{S}_i \propto \rho_0$, the hierarchy between the finite profiles at large $\lambda$ is manifest. The tophat profile, enclosing the most mass for the same volume within $R$, exhibits the largest signal, whereas the exponential profile, enclosing the least, leads to the smallest signal. Moreover, since $b < R$, only in the point-like scenario does the sensor see the whole DM mass over the entire sensor trajectory, resulting in the strongest signal for the point-like profile. 

Conversely, when $\lambda \ll b$, the signal for the finite density profiles scale as $\lambda^2$ instead of the exponential suppression we see from the point-like result. This important difference can be seen by considering the Helmholtz equation that the Yukawa potential satisfies:
\begin{equation}
    \left(\nabla^2 - \frac{1}{\lambda^2}\right)U(\boldvec{r}) = 4 \pi \alpha G \rho(\boldvec{r}) \implies U(\boldvec{r}) \simeq - 4 \pi \alpha G \lambda^2 \rho(\boldvec{r}) + \mathcal{O}(\lambda^4)\,.
\end{equation}
This result is valid when the density profile varies slowly and smoothly over the scale of the screening length $\lambda$: $\lambda^2 |\nabla^2\rho| / \rho \ll 1$. Taking the characteristic scale over which a given density profile varies to be $R$, such that $|\nabla^2\rho| \sim \rho / R^2$, then we indeed see that this is true for $\lambda / R \ll 1$.  The Yukawa force is then $F(\boldvec{r}) \simeq 4 \pi \alpha G \lambda^2 \boldvec{\nabla}\rho(\boldvec{r})$, leading to the power-law behavior in the signal. Thus, the clump signal dominates over the point-like result for small enough $\lambda$.

From \cref{fig:single_sig}, we see that we expect the largest signal to arise when $\lambda \rightarrow \infty$. Since the point-like result dominates in this regime, it is instructive to consider the point-like case to gain further intuition on the signal size and therefore the ultimate sensitivity that an array can have. In this case, the impulse delivered to a sensor will be due to the entire mass of the DM clump. \cref{eq:sig} can then be evaluated analytically, for which we find
\begin{equation}\label{eq:s_approx}
    \mathcal{S}_i = (1 + \alpha) \frac{2 G M m_s}{b v} \simeq 1\times 10^{-18}\,\mrm{N\,s}\,\bigg(\frac{\alpha}{10^4}\bigg) \left(\frac{M}{\MP}\right) \left(\frac{m_s}{100\,\mrm{g}}\right)\left(\frac{1\,\mrm{cm}}{b}\right)\,,
\end{equation}
where we have fiducialized to a value of $\alpha \gg 1$ that we will ultimately be sensitive to at these large screening lengths.

\subsubsection{Impulse Noise of a Single Sensor}
\label{subsubsec:noise}

When searching for the small impulses imparted on a sensor, we are continuously monitoring its center-of-mass displacement, attempting to detect motions above a noise floor. This noise is due to both the inherent coupling between the sensor and the environment, leading to thermal and seismic noise, as well as to the particular scheme used to measure these displacements, resulting in quantum noise~\cite{Caves:1980NI1,Caves:1981hw,Hadjar:1999vx}. Our goal is to minimize these noise terms as much as possible to maximize our signal-to-noise ratio (SNR), therefore optimizing our sensitivity. Ultimately, if a signal were observed, this prescription would give us the greatest inferencing power in discerning the properties of the clump and the nature of the force between the clump and the sensor.
 
Thermal noise arises from the sensor's interactions with its finite-temperature environment.  
These interactions cause random fluctuations in the sensor’s motion due to the stochastic energy exchange with the surroundings. 
The behavior of this noise is well described by the fluctuation–dissipation theorem, which quantitatively relates the dissipative properties of the system to its thermal fluctuations. Assuming that measurements are made at regular intervals separated by a duration $\tau$, then the root-mean-square thermal noise is calculated as~\cite{Qin:2025jun}
\begin{align}\label{eq:Nthermal}
    \mathcal{N}_\mrm{th} = \sqrt{\alpha_\mrm{therm} \tau}\,, 
    \qquad \text{where} \qquad 
    \alpha_\mrm{therm} \equiv 4 m_s k_B T \gamma\,.
\end{align}
Here, $T$ is the temperature of the surroundings and $\gamma$ is the mechanical damping rate, which dictates the rate at which energy is lost to the environment. As expected from a Brownian process, the thermal noise grows with a characteristic $\tau^{1/2}$ scaling.

On the other hand, quantum noise is intrinsic to the quantum nature of a measurement
itself, arising from the inherent fluctuations in the signal (shot noise) and the back action experienced by the system by the act of performing a measurement. This noise term is governed by the Heisenberg uncertainty principle, such that a more precise position measurement implies a larger momentum uncertainty. 
Consequently, a higher sampling rate, corresponding to smaller values of $\tau$, results in higher quantum noise.  If measurements are made at regular intervals of duration $\tau$, then the root-mean-square quantum noise can be written as~\cite{Qin:2025jun}
\begin{align}\label{eq:Nquantum}
    \mathcal{N}_\mrm{quant} = \sqrt{\frac{\beta_\mrm{quant}}{\tau}} \,,
    \qquad \text{where} \qquad 
    \beta_\mrm{quant} \equiv \frac{4 m_s \hbar}{\xi^2}
    \,.
\end{align}
The factor $\xi$ is the quantum noise compression factor. When $\xi = 1$, the quantum noise is said to be at the level of the `standard quantum limit', whereas if $\xi > 1$, the system has been subjected to quantum noise reduction techniques, pushing the quantum noise below this limit. For momentum sensing with mechanical oscillators, both squeezed light readouts (for optomechanical sensors) and back-action evading techniques are able to achieve this \cite{Ghosh:2019rsc,Lee:2025hkp}.  

Assuming that the thermal and quantum noise terms are the dominant noise sources and that they are uncorrelated to one another, the total noise $\mathcal{N}$ is calculated by summing these contributions in quadrature: 
\begin{equation}\label{eq:N}
\begin{split}
    \mathcal{N} &= \sqrt{\mathcal{N}_\mrm{therm}^2 + \mathcal{N}_\mrm{quant}^2} \geq \left(4 \alpha_\mrm{therm} \beta_\mrm{quant}\right)^{1/4} \\
    &\simeq 6 \times 10^{-19}\,\mrm{N\,s}\,
    \left(\frac{30}{\xi}\right)^{1/2}\left(\frac{m_s}{100\,\mrm{g}}\right)^{1/2}\left(\frac{T}{15\,\mrm{mK}}\right)^{1/4}\left(\frac{\gamma}{10^{-11}\,\mrm{Hz}}\right)^{1/4}\,,
\end{split}
\end{equation}
where we have fiducialized to our detector parameters, further described below. The lower bound on this noise is due to the fact that the thermal noise component grows with larger measurement times $\tau$, while the quantum noise increases for smaller times. The total noise is minimized when $\tau = \tau_\mrm{opt} \equiv \sqrt{\beta_\mrm{quant} / \alpha_\mrm{therm}}$. For the fiducial parameters introduced above, this evaluates to $\tau_\mrm{opt} \approx 0.24 \, \mathrm{s}$. 
As we remarked already above, for the parameters of interest in this work, we find $t_\mrm{imp} \ll \tau_\mrm{opt} \ll 2\pi/\omega_0 \ll 1/\gamma$, and we therefore choose $\tau = \tau_\mrm{opt}$. See also \rref{Qin:2025jun} for a discussion of other scenarios.

Beyond thermal and quantum noise, vibration noise can also arise from various sources. This includes including seismic activity both within and outside of the cryogenic system. We do not include vibrational noise in   this study, as sufficient vibration isolation to ensure that a levitated particle is dominated by thermal noise at $20\,\mathrm{mK}$ has been demonstrated at a resonance frequency of $1\,\mathrm{Hz}$ at ground level~\cite{vanHeck:2022evk}. 
In addition, it is possible to improve vibration isolation further by situating the experiment in an underground laboratory, where baseline vibrational noise is approximately two orders of magnitude lower than on the surface of an urban area~\cite{Trozzo:2022tar}.

\subsubsection{Signal-to-Noise Ratio of a Sensor Array}
\label{subsubsec:snr}

Using \cref{eq:sig,eq:N}, we can define the signal-to-noise ratio (SNR) for a given sensor:
\begin{align}\label{eq:SNRi}
    \mathrm{SNR}_{i} \equiv \frac{\mathcal{S}_{i}}{\mathcal{N}} 
    \,,
\end{align}
where signal $\mathcal{S}_{i}$ must be calculated for a particular sensor trajectory. 
Assuming that the noise between each sensor is uncorrelated, the total SNR over the entire array is given by summing the individual sensor SNRs in quadrature, 
\begin{align}\label{eq:SNR}
    \mathrm{SNR} \equiv \sqrt{ \sum_{i=1}^{N^3} \biggl( \mathrm{SNR}_{i} \biggr)^2 }
    \,.
\end{align}

As one final estimate, let us consider the total SNR of an array composed of an even number of sensors along a single dimension $N$. For simplicity, let us assume  that a DM clump passes directly through the center of the array and that the entire array moves along the $z$-direction. The SNR will be dominated by the sensors that are closest to the clump, resulting in a closest approach distance for any single sensor of $b \sim d / \sqrt{2} \approx 7\,\mrm{cm}$. Along any row of sensors, there will be $N$ equivalent sensor SNRs, and for our even setup there will be $4$ such copies of these. The total SNR is then $\mrm{SNR} \simeq \sqrt{4 N\, (\mrm{SNR}_i)^2} = 2 \sqrt{N}\, \mrm{SNR}_i$. Moreover, the projection of the impulse onto a sensitivity axis pointing along $\hat{\bm{n}} = \hat{\bm{x}}$ will yield a $1 / \sqrt{2}$ suppression. Inserting this into our estimate of the single-sensor signal of \cref{eq:s_approx}, we find that
\begin{equation}
\begin{split}
    \mrm{SNR} 
    \simeq&~3\, \bigg(\frac{\alpha}{10^4}\bigg) \left(\frac{M}{\MP}\right) \left(\frac{7\,\mrm{cm}}{b}\right) \\
    &\times \left(\frac{N}{20}\right)^{1/2}\left(\frac{m_s}{100\,\mrm{g}}\right)^{1/2}\left(\frac{\xi}{30}\right)^{1/2}\left(\frac{15\,\mrm{mK}}{T}\right)^{1/4} \left(\frac{\gamma}{10^{-11}\,\mrm{Hz}}\right)^{1/4}
    \,.
\end{split}
\end{equation}
From this estimate, we expect that an experiment with these parameters  will be sensitive to interaction strengths $\alpha \gtrsim 10^{4}$ for a DM clump mass of $M \approx \MP$ at large values for the screening length $\lambda$.

\subsection{Numerical Simulation}
\label{subsec:sim}


\begin{table}[p]
\renewcommand{\arraystretch}{1.5}
    \centering
\begin{tabular*}{1\columnwidth}{@{\extracolsep{\fill}}lr}
    \toprule\midrule
    Parameter & Value \\
    \midrule
    Dark Matter Density Profile ($\rho$) & Point-like, Tophat, Gaussian, Exponential\\
    Dark Matter Mass ($M$) & $10^{-1}\,\MP\text{~to~}10\,\MP$\\
    Dark Matter Radius ($R$) & $10^{-3}\,\mrm{m}\text{~to~}10\,\mrm{m}$\\
    Yukawa Strength ($\alpha$) & $10^{4} \text{~to~}10^{8}$\\
    Yukawa Screening Length ($\lambda$) & $10^{-3}\,\mrm{m}\text{~to~} 10^2\,\mrm{m}$\\
    \midrule[0.25pt]
    \midrule[0.25pt]
    Array Dimensions ($N \times N \times N$) & $20 \times 20 \times 20$\\
    Sensor Spacing ($d$) & $10\,\mrm{cm}$\\
    Array Length ($L$) & $d(N-1) = 1.9\,\mrm{m}$\\
    Sensor Mass ($m_s$) & $100\,\mrm{g}$\\
    Sensor Density ($\rho_s$) & $\rho_\mrm{Pb} \approx 11.34\,\mrm{g\,cm^{-3}}$\\
    Sensor Radius ($R_s$) & $[3 m_s / (4 \pi \rho_s)]^{1/3} \approx 1.3\,\mrm{cm}$\\
    Array Temperature ($T$) & $15\,\mrm{mK}$\\
    Resonant Frequency ($\omega_0$) & $2 \pi (20\,\mrm{m Hz})$\\
    Quality Factor ($Q$) & $10^{10}$\\
    Damping Rate ($\gamma$) & $\omega_0 / Q \approx 1.3\times10^{-11}\,\mrm{Hz}$\\
    Quantum Noise Squeezing ($\xi$) & $15\,\mrm{dB}$\\
    Measurement Time ($\tau$) & $0.24\,\mrm{s}$\\
    Exposure Time ($t_\mrm{exp}$) & $5\,\mrm{yr}$\\
     \midrule
     \bottomrule 
\end{tabular*}
\caption{The theoretical (top) and experimental (bottom of table) parameters we consider in this work. For the theoretical parameters, included are the typical values they take in our analysis. The functional forms for the dark matter density profiles are given in \cref{eq:rho_profs}.}
\label{tab:params}
\end{table}

To estimate the total expected SNR of an array to a passing DM clump, we perform a series of Monte Carlo simulations. For a given choice of the DM density profile, this measure allows us to draw sensitivity projections for the various theoretical parameters that control the signal: the dark matter mass, the characteristic radius of a DM clump, the Yukawa strength parameter, and the Yukawa force screening length. These are summarized in \cref{tab:params}, where we also give the range of values for which we run our analysis. Beyond these variables, we also assume a particular set of experimental parameters to perform our simulations, which we expand on below.

For our experimental configuration, we assume the `future milestone' benchmark described in \rref{Qin:2025jun} and summarized in \cref{tab:params}. Featuring a $20\times20\times20$ array of $m_s = 100\,\mrm{g}$ lead sensors, the setup is cooled to $T = 15\,\mrm{mK}$ and employs a quantum noise reduction level of $\xi = 15\,\mrm{dB}$. The resonant angular frequency of the sensors is set to $\omega_0 = 2\pi(20\,\mrm{m Hz})$, and the quality factor is taken to be $Q = 10^{10}$. Taken together, these two quantities define the damping rate $\gamma \equiv \omega_0 / Q \approx 1.3\times10^{-11}\,\mrm{Hz}$. 
This setup is inspired by a future realization of a mechanical sensing array composed of magnetically levitated particles~\cite{Latorre:2022vxo,Vinante:2020zjt,Hofer:2022chf,Fuchs:2023ajk,Schmidt:2024hdc}. Magnetic levitation technology makes for highly sensitive force sensors, affording us low thermal noise and heavy test masses~\cite{Vinante:2020zjt,Fuchs:2023ajk,Janse:2024kcn}. This latter feature gives us greater sensitivity to the fifth force we consider since it couples proportionally to mass. For the lattice dimensions, we choose a sensor spacing of $d = 10\,\mrm{cm}$, corresponding to a total array length of $L = d(N-1) = 1.9\,\mrm{m}$.

Assuming this experimental realization, we implement our numerical simulations following a similar procedure to that of \cite{Qin:2025jun}, summarized below and expanded on in \cref{sec:num_alg}. For each choice of DM density profile, we begin by selecting our theory parameters, the ranges for which are given in \cref{tab:params}. We then initialize our cubic sensor array far away from the DM clump with a random orientation and a speed drawn from a Maxwell-Boltzmann distribution, taking  $v_\mrm{rms} = v_0 / \sqrt{2}$ with $v_0 = 238\,\mrm{km\,s^{-1}}$. We compute the force on any given sensor along its trajectory using \cref{eq:Fvec}, from which we compute the total SNR via \cref{eq:SNR}. 

We repeat the above procedure for a total of $10^3$ potential tracks through the array, recording the total SNR in each run. From the distribution of SNRs, we find the probability $\mathbb{P}_{10}$ of detecting an SNR greater than $10$. This high requirement for the SNR is imposed by the look-elsewhere effect, and it is the threshold suggested by \rref{Qin:2023urf} for a search of duration $\usim1\,\mathrm{yr}$ or greater. Finally, we project our sensitivity to a particular model parameter by requiring that $N_\mrm{encounter} \mathbb{P}_{10} \geq 1$, where
\begin{equation}\label{eq:N_encounter}
    N_\mrm{encounter} \equiv \mathcal{R} t_\mrm{exp}\simeq  1 \left(\frac{\MP}{M}\right)\left(\frac{A}{1\,\mrm{m^2}}\right) \left(\frac{t_\mrm{exp}}{5\,\mrm{yr}}\right)
\end{equation}
is the expected number of DM encounters with the array. Here,  $t_\mrm{exp}$ is the experimental exposure time, $\mathcal{R}$ is the DM rate given in \cref{eq:rate}, and $A \equiv \pi(R + L/2)^2$ is a geometrical area that accounts for the finite size of the DM clump. Satisfying this inequality ensures that we are likely to observe at least one DM event with an SNR of at least $10$ over the duration of the experiment.

Before continuing to our results, we note that fixing our lattice spacing $d$ results in a typical impact parameter $\langle b \rangle$ for a straight-line trajectory passing through the array when considering many randomized tracks. This value is parametrically smaller than $d$ because of projection effects. Concretely, consider a cubic lattice with $N^3$ vertices spread over a volume $V = L^3$, where $L = (N-1)d$.  If the lattice is projected onto a plane, then the surface density is $\sigma \sim N^3 / L^2$, and the typical separation between points is $\sigma^{-1/2} \sim L / N^{3/2} = (N^{-1/2} - N^{-3/2}) d$.  The typical impact parameter can be estimated as half of this quantity, which to leading order is given by $\langle b \rangle \simeq d/(2\sqrt{N}) \approx 1~\mrm{cm}$.

\section{Projected Sensitivities}
\label{sec:results}

For a given choice of DM density profile, the total SNR measured by an array is controlled by each of our four theory parameters. These are the force strength parameter $\alpha$, the Yukawa screening length $\lambda$, the mass of the DM clump $M$, and its characteristic radius $R$. Due to the high dimensionality of this parameter space, we predominantly explore our sensitivities in a selection of 2D slices through this space, holding the remaining model parameters fixed. We consider the total signal $\mathcal{S}_\mrm{tot}$ across all sensors in an array with $R$, as well as the $M\text{--}R$, $\lambda\text{--}R$, $\alpha\text{--}\lambda$, and $\alpha\text{--}M$ planes.

\subsection{Total Signal with Dark Matter Radius}
\label{subsec:impulse}

\begin{figure}[t]
    \centering
    \includegraphics{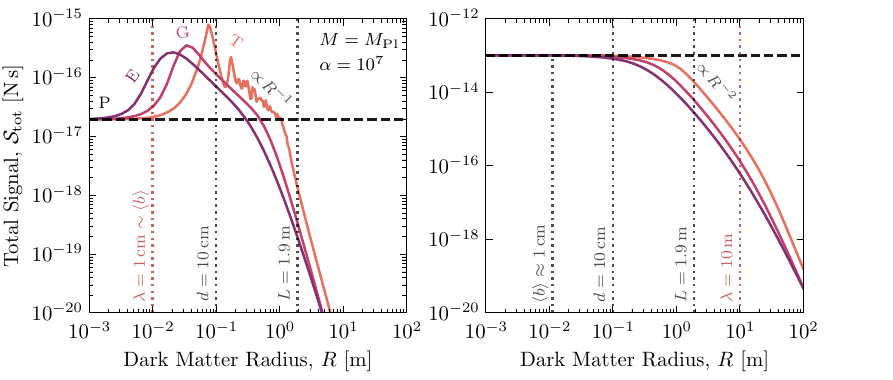} 
    \caption{\label{fig:impulse_R}
    The summed signal (cf.~\cref{eq:sig}) with characteristic dark matter clump radius $R$ for the different dark matter density profiles $\rho$ we consider in this work (cf.~\cref{eq:rho_profs}). The vertical lines indicate the geometrical scales of the detector array, including  the sensor spacing $d = 10\,\mrm{cm}$ and the array length $L = 1.9\,\mrm{m}$, as well as the expected value of the impact parameter $\langle b \rangle = d / (2 \sqrt{N}) \approx 1\,\mrm{cm}$, with $N = 20$ the number of sensors along a single array dimension. Results are computed for dark matter mass $M = \MP$ and strength parameter $\alpha = 10^7$. The signal scales as $\mathcal{S}_\mrm{tot} \propto \alpha M$ if the Yukawa force dominates over the gravitational force. \textbf{Left:} The result for $\lambda = 1\,\mrm{cm}$. For radii $\lambda \lesssim R \lesssim L$, the signal goes as $R^{-1}$. \textbf{Right:} The result for the screening length $\lambda = 10\,\mrm{m}$. For radii $d \lesssim R \lesssim \lambda$, the signal goes as $R^{-2}$. For both choices of $\lambda$, the relationship of the signal to $R$ depends on the assumed density profile at large DM radii.} 
\end{figure}

Before exploring our sensitivities, it is first instructive to consider a simpler issue: the total signal over the array. This serves to illustrate how these parameters can interplay with one another in a non-trivial way and how the detector scales impact the ultimate signal. Assuming the detector configuration outlined in \cref{tab:params}, we send the sensor array through the center of a DM clump, computing the signal $\mathcal{S}_i$ for each sensor using \cref{eq:sig} and then summing them to find the total signal, defined as $\mathcal{S}_\mrm{tot} \equiv\sum_{i=1}^{N^3} \mathcal{S}_i$.

In \cref{fig:impulse_R}, we show how this signal varies with DM clump radius $R$ for a DM clump of mass $M = \MP$ and Yukawa interaction strength of $\alpha = 10^{7}$. To highlight how different choices for the screening length $\lambda$ can lead to unique behaviors cross the experimental scales of the array, we compute the signal for two values of $\lambda$. Our first choice, $\lambda = 1\,\mrm{cm}$, is chosen to be smaller than the smallest scale of array: the detector spacing $d = 10\,\mrm{cm}$. Our second choice, $\lambda = 10\,\mrm{m}$, is instead picked to be larger than the largest experimental scale: the array length $L \approx 1.9\,\mrm{m}$. 

The signal behavior for small $\lambda$ is shown in the left panel of \cref{fig:impulse_R}, for which $\lambda = 1\,\mrm{cm} \sim \langle b \rangle \ll d$. When $\lambda + R \ll \langle b\rangle$, 
the fifth force is completely screened before reaching a sensor, and the interaction is purely gravitational. As $\lambda + R$ increases to and beyond $\langle b\rangle$, the fifth force begins to grow as less screening occurs. However, for a fixed DM clump mass, increasing the characteristic radius beyond a certain threshold leads to a decrease in the signal since $\mathcal{S}_i \propto \rho_0 \sim M / R^3$, as we argued in \cref{subsec:signal} and is demonstrated in the right panel of \cref{fig:single_sig}. Then, for the total array signal, the number of sensors $N_s$ piercing through the cylinder traced out by the trajectory of the clump is  $N_s \propto R^2 L$. Therefore, the total signal scales as $\mathcal{S}_\mrm{tot} \sim N_s \mathcal{S}_i \propto R^{-1}$. Beyond $R \gtrsim L$, the scaling relation is dependent on the density profile, roughly following $\mathcal{S}_\mrm{tot} \propto R^{-4}$. Note that the behavior of the tophat profile is due to a combination of the jump discontinuity in this profile and the DM clump gradually absorbing more sensors in the array as it increases in size.  At intermediate values of $R$, the signal from the extended DM clump surpasses that from the point-like result due to an interplay between the number of sensors being swallowed by the clump and the power-law vs exponential signal suppression between these models.

The large $\lambda$ behavior of the total signal is shown in the right panel of \cref{fig:impulse_R}, for which $\lambda = 10\,\mrm{m} \gg L$. In this case, when $R\ll L$, the Yukawa force acts as a coherent, long-range force over the entire array, producing a constant signal that is independent of the DM scale $R$. However, as $R$ grows past $L$, the sensors are enclosed by decreasing mass as the density falls for a constant DM clump mass, similar to the argument given above for small $\lambda$. In this case, we have a similar scenario to the large $\lambda$ regime for the single sensor signal, shown in \cref{fig:single_sig}. This explains the $R^{-2}$ behavior. For larger $R$, the signal is dependent on particular DM clump profile assumed, but it roughly follows  $\mathcal{S}_\mrm{tot} \propto R^{-4}$.

\begin{figure}[t]
    \centering
    \includegraphics{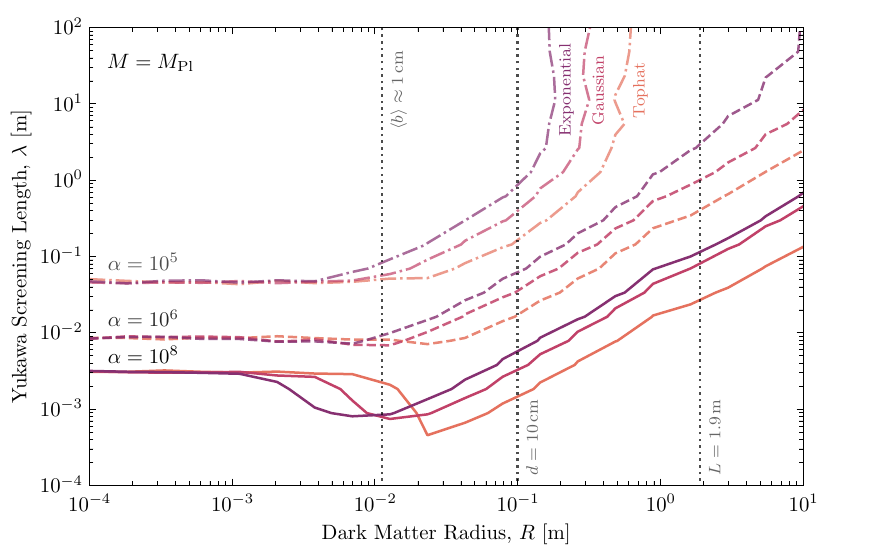}
    \caption{\label{fig:lambda_R}
    Projected sensitivities to the Yukawa screening length $\lambda$ with characteristic dark matter clump radius $R$ for the dark matter density profiles we consider in this work (cf.~\cref{eq:rho_profs}). Sensitivities are drawn for strength parameters $\alpha \in \{10^{5}, 10^{6}, 10^{8}\}$. The vertical lines indicate the geometrical scales of the detector array, including  the sensor spacing $d = 10\,\mrm{cm}$ and the array length $L = 1.9\,\mrm{m}$, as well as the expected value of the impact parameter $\langle b \rangle = d / (2 \sqrt{N}) \approx 1\,\mrm{cm}$, with $N = 20$ the number of sensors along a single array dimension. Results are computed for dark matter mass $M = \MP$.}
\end{figure}

\subsection{Sensitivity in the $\lambda\text{--}R$ plane}
\label{subsec:sensitive_lam_R}

In \cref{fig:lambda_R}, we show a two-dimensional slice of our parameter space in which we vary the Yukawa screening length $\lambda$ and the characteristic radius of the DM clump $R$.  
We fix the dark matter clump mass to $M = \MP$, and we show three values of the Yukawa strength $\alpha$ for each of the three density profiles. 
Regions above the curves, typically larger $\lambda$ and smaller $R$, are projected to be within reach of the experiment.  

For models with small clump radius $R \lesssim \langle b \rangle \approx 1\,\mrm{cm}$, the projected sensitivity becomes insensitive to $\lambda$ and the density profile $\rho$, converging to the point-like result when $R \lesssim 10^{-3}\,\mrm{m}$.  
In this limit, the dark matter clump is effectively point-like, so its radius and density profile are irrelevant for detection.  
The experiment would be sensitive to models with $\lambda \sim \langle b \rangle \approx 10^{-2}\,\mrm{m}$, with a weak logarithmic dependence on $\alpha$. If $R + \lambda \lesssim \langle b \rangle$, then the dark matter clump is small enough and the range of its Yukawa force is so short that the closest sensor to the DM clump will feel an exponentially weak force $\propto \alpha \, \ee^{-\langle b \rangle / \lambda}$.  
For exponentially-larger values of $\alpha$, it is possible to probe smaller $\lambda$.  

For models with $R \gtrsim \langle b \rangle \approx 1\,\mrm{cm}$ the clump is large enough to envelope at least a few sensors.  
Further increasing $R$ tends to reduce the force felt by any individual sensor due to the decreasing density for a fixed $M$, and the sensitivity degrades; see the $\lambda \propto R^1$ scaling.  
Order-one differences between the three density profiles develop based on their spatial extent and the amount of mass contained within the same volume.  

For models with $\alpha = 10^5$ and $\lambda \gtrsim \mathrm{few} \times \mrm{m}$, the sensitivity curves turn upward.  
In this region, the force is effectively long-range, with the Yukawa screening length becoming larger than the size of the detector array.  
The signal then becomes independent of $\lambda$.  
Increasing $R$ further degrades the signal by diluting the force across a larger volume and reducing the force felt by any individual sensor. All curves suffer this suppression at large enough $R$ since, as argued in \cref{subsec:signal} and also observed in \cref{subsec:impulse}, the total signal for large $\lambda$ scales as $\mathcal{S}_i \propto \alpha  R^{-2}$. Therefore, if we are at the threshold of the signal required for a detection, increasing $\alpha$ by an order of magnitude will tend to delay this suppression in $R$ by two orders of magnitude.

It is interesting that the signal does not degrade monotonically for models with increasingly large dark matter clump radius $R$.  
Rather, for the models with large Yukawa strength, see the $\alpha = 10^8$ curves, the best sensitivity is achieved when the dark matter clump radius is comparable to the typical impact parameter for the array, $R \approx \langle b \rangle$.  
Here the dark matter clump is just large enough to envelop a few sensors but not so large that its force is diluted across many sensors. This increased sensitivity over the point-like result in the small $\lambda$ regime is due to the power law behavior of the clump signal over the exponential scaling in the point-like case.
The improvement in sensitivity at $\langle b \rangle \approx R$ could be used to design detectors that target specific models of composite UHDM.  

\subsection{Sensitivity in the $M\text{--}R$ plane}
\label{subsec:sensitive_M_R}

\begin{figure}[t]
    \centering
    \includegraphics{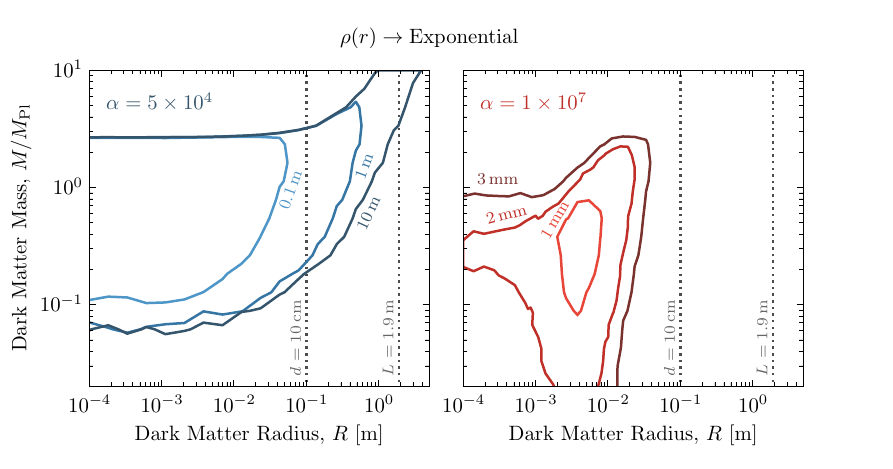}
    \caption{\label{fig:M_R}
    Projected sensitivities to the dark matter clump mass $M$ with characteristic dark matter clump radius $R$ for the exponential density profile (cf.~\cref{eq:rho_profs}). The contours are drawn for the shown choices of the Yukawa screening length $\lambda$. An experiment has sensitivity inside the contours. The vertical lines indicate the geometrical scales of the detector array, including  the sensor spacing $d = 10\,\mrm{cm}$ and the array length $L = 1.9\,\mrm{m}$. \textbf{Left:} Sensitivities  for the strength parameter $\alpha = 5\times 10^{4}$. \textbf{Right:} Sensitivities  for the strength parameter $\alpha = 1\times10^{7}$. 
    }
\end{figure}

In \cref{fig:M_R}, we show a two-dimensional slice of our parameter space in which we vary the dark matter clump's mass $M$ and scale radius $R$.  
We show two values of the Yukawa strength $\alpha$ and several values of the Yukawa screening length $\lambda$. 
For this illustration, we only consider the exponential density profile.  
The ``interior'' regions bounded by the curves, typically at smaller $R$, are projected to be within reach of the experiment.  

Towards large values of $M$, the sensitivity degrades because of flux considerations.  
See the discussion at \cref{eq:rate}.  
For models with larger $M$, the local number density of dark matter must be smaller than $n_\mrm{DM} \propto M^{-1}$ to ensure that the local mass density of dark matter agrees with the observed value.  
Any direct detection experiment loses sensitivity to large $M$ because the expected number of dark matter particles that encounter the detector drops below order one.  
For our fiducial size and exposure, this threshold is $M \approx 3 \MPl$ at small $R$.  
For larger $R \gtrsim L$, this threshold weakens as $M \propto R^2$ since the clump's size sets the area for the flux calculation.  

Towards small values of $M$, the sensitivity degrades because of force considerations.  
We parametrize the gravitational and Yukawa forces to be proportional to the sensor and dark matter clump masses---see \cref{eq:potential_energy}.  
For models with smaller $M$ and fixed $\alpha$, the force due to the dark matter clump is weaker. 
For $\alpha = 5 \times 10^4$, we find that sensitivity is lost for masses below $M \sim 10^{-1} \MPl$.  

The two panels show different values of the Yukawa interaction strength $\alpha$.  
For smaller $\alpha$, the sensitivity becomes insensitive to $R$ for small values of $R$, and for larger $\alpha$ the sensitivity is largest at intermediate values.  
This behavior can also be seen by comparing with \cref{fig:lambda_R}.  

\subsection{Sensitivity in the $\alpha\text{--}\lambda$ plane}
\label{subsec:sensitive_alpha_lambda}

\begin{figure}[t]
    \centering  
    \includegraphics[]{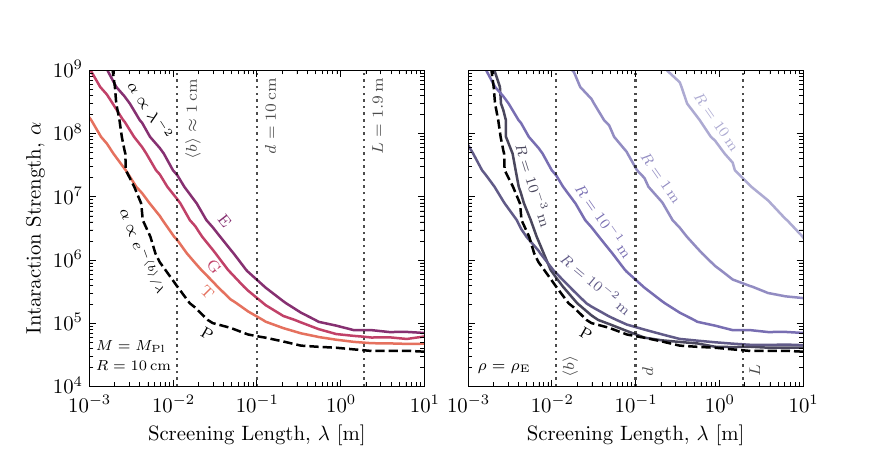}
    \caption{\label{fig:alpha_lambda}
    Projected sensitivities to the Yukawa interaction strength $\alpha$ with varying screening length $\lambda$ for fixed DM clump mass $M = \MP$. The vertical lines indicate the geometrical scales of the detector array, including  the sensor spacing $d = 10\,\mrm{cm}$ and the array length $L = 1.9\,\mrm{m}$, as well as the expected value of the impact parameter $\langle b \rangle = d / (2 \sqrt{N}) \approx 1\,\mrm{cm}$, with $N = 20$ the number of sensors along a single array dimension. \textbf{Left:} Sensitivities for the density profiles we consider in this work (cf.~\cref{eq:rho_profs}) for fixed DM clump radius $R = 10\,\mrm{cm}$. For $\lambda \ll R \sim \langle b \rangle$, the sensitivities follow a power law scaling $\alpha \propto \lambda^{-2}$; this is to be contrasted with the point-like result, where $\alpha \propto e^{-\langle b \rangle / \lambda}$. \textbf{Right:} Sensitivities for the exponential density profile for varying DM clump radius. For $R \lesssim \langle b \rangle$, the curves converge to the point-like result.
    }
\end{figure}

In \cref{fig:alpha_lambda}, we show a two-dimensional slice of our parameter space in which we vary the Yukawa interaction strength $\alpha$ and the screening length $\lambda$. 
We hold fixed the dark matter clump's mass $M = \MPl$, and we show several values of the clump's radius $R$ for different density profiles.  
Regions above the curves, toward larger $\alpha$ and $\lambda$, are projected to be within reach of the experiment.

For large values of the Yukawa screening length $\lambda$, the sensitivity becomes insensitive to $\lambda$---notice how the curves flatten above $\lambda \approx 10^0~\mrm{m}$.  
This occurs because $\lambda \gtrsim L = 1.9~\mrm{m}$, and the force is effectively long-range such that the entire detector array can be fit within the Coulomb-like part of the Yukawa potential.  
In this long-range regime, the projected sensitivity is approximately $\alpha \approx \mathrm{few} \times 10^{4}$ for the range of $R$ and density profiles shown. 
This result agrees well with the results of \rref{Qin:2025jun}, which found $\alpha \sim 6 \times 10^4$ for the same detector array configuration that we study (also see below).

A key difference between the point-like and extended cases is how the sensitivity to the interaction strength $\alpha$ degrades with decreasing screening length $\lambda$. For a point-like DM particle, the sensitivity to $\alpha$ falls off exponentially due to the exponential screening of the Yukawa force. However, this fall off follows a different relationship for an extended DM candidate when its characteristic radius $R$ increases beyond the expected impact parameter $\langle b \rangle \sim 1\,\mrm{cm}$.  As we argued in \cref{subsec:signal}, the total signal over the array scales as $\mathcal{S} \propto \alpha\lambda^2$ when $R \gtrsim \langle b \rangle$, such that we require $\alpha \propto \lambda^{-2}$ to produce the same SNR. Thus, for DM candidates with greater spatial extent than approximately the spacing in a sensor array, the sensitivity drops off only as a power law in $\lambda$ instead of being exponentially suppressed. This is an important phenomenological distinction, implying that such extended candidates would be significantly easier to detect compared to point-like DM particles when the screening length of the force is small compared to the inter-sensor spacing.

Searches for a fifth force that mediates interactions among visible-sector matter are complementary to the projections that we provide in this section.  
If the force is modeled as a Yukawa interaction and if visible-sector matter has a charge-to-mass ratio on the order of $Q_v/M_v = 1\,\mathrm{GeV}^{-1}$, then the strength of the fifth force should be below $O(0.1\%)$ of gravity for $\lambda \sim 1\,\mrm{mm}\text{--}1\, \mrm{m}$, where torsion balances are most sensitive~\cite{Salumbides:2013dua}.  
If one also assumes a charge-to-mass ratio for the dark sector matter $Q_d/M_d$, then the fifth force limits can be recast as upper limit on the visible-dark interaction strength $\alpha$, which is the parameter that we probe.  
However, for reasonable values of $Q_d/M_d$, the recast limits are not constraining. 
Generically, we may write our coupling $\alpha$ in terms of the visible and dark sector couplings $\alpha_v$ and $\alpha_d$ as
\begin{equation}
    \alpha = \frac{\alpha_v \alpha_d
    }{4 \pi G}\left(\frac{Q_v}{M_v}\right) \left(\frac{Q_d}{M_d}\right)\,.
    \label{eq:matched_couplings}
\end{equation}
Mapping to any particular $\alpha_v$ or $\alpha_d$ on which we may already have constraints then requires us to assume a particular model for how the DM can interact with us.

For instance, the authors of \rref{Qin:2025jun} also studied the prospects of detecting point-like ultraheavy dark matter that couples to an accelerometer through a long-range fifth force interaction.  Writing the force law as $F = \tilde{\alpha} N_\mrm{nuclei} / r^2$, where $N_\mrm{nuclei}$ is the number of nuclei in the test mass, they calculated the projected sensitivity to the parameter $\tilde{\alpha}$. Thus, in this model, we have that $\alpha_v = \tilde{\alpha}$, $Q_v = N_\mrm{nuclei}$, and $\alpha_d Q_d = 4\pi$.  For their ``future milestones'' design configuration, they found that couplings larger than $\tilde{\alpha} \sim 10^{-12}$ could be probed for dark matter masses up to $M \approx 3 \times 10^{19} \, \mathrm{GeV}$ (cf., Fig.~4, ``Future Milestone'').   Then $Q_v / M_v \sim N_\mrm{nuclei}/m_s \sim 1 / m_\mrm{nucleus}$, which approximates the sensor's mass as $m_s = m_\mrm{nucleus} N_\mrm{nuclei}$, for which we take $m_\mrm{nucleus} = 207 \, \mathrm{GeV}$ (lead) in this estimate. Using \cref{eq:matched_couplings}, we therefore find that $\alpha \sim 6 \times 10^4$ in our convention. Moreover, since our configuration parallels their `future' setup, which itself probes couplings $\tilde{\alpha}$ many orders of magnitudes smaller than the present leading bounds from liquid xenon direct detection experiments, we find the same result, where all of our limits surpass the leading direct detection limits. More broadly, our results could be mapped to any particular model assumption via \cref{eq:matched_couplings}.


\subsection{Sensitivity in the $\alpha\text{--}M$ plane}
\label{subsec:sensitive_alpha_M}

\begin{figure}[t]
    \centering  
    \includegraphics{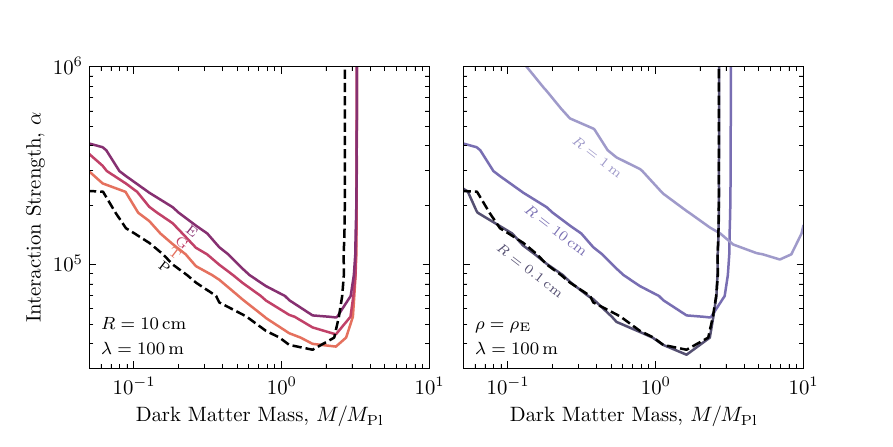} 
    \caption{\label{fig:alpha_M}  
    Projected sensitivities to the Yukawa interaction strength $\alpha$ with varying dark matter clump mass. \textbf{Left:} Sensitivities for the density profiles we consider in this work (cf.~\cref{eq:rho_profs}) for a fixed DM clump radius $R = 10\,\mrm{cm}$ and Yukawa screening length of $\lambda = 100\,\mrm{m}$. \textbf{Right:} Sensitivities for the exponential density profile for varying clump scale radius $R$ and a fixed Yukawa screening length of $\lambda = 100\,\mrm{m}$. 
    }
\end{figure}

In \cref{fig:alpha_M}, we show a two-dimensional slice of our parameter space in which we vary the Yukawa interaction strength $\alpha$ and dark matter clump's mass $M$. 
We show our sensitivities using all of our considered density profiles for a fixed scale radius $R = 10\,\mrm{cm}$ and Yukawa screening length $\lambda = 100\,\mrm{m}$. We also show the result for the exponential profile for different values of the clump radius, showing the convergence to the point-like result for small $R$.  
Regions above the curves (larger $\alpha$) are projected to be within the experiment's reach.

For large values of $M$ there is a sharp drop in sensitivity---a feature also present in \cref{fig:M_R}. This cutoff occurs when the inequality $N_\mrm{encounter} \mathbb{P}_{10} \geq 1$ can no longer be satisfied.  
Large $M$ implies small $N_\mrm{encounter}$ as in \cref{eq:N_encounter}, and if $N_\mrm{encounter} < 1$, then $N_\mrm{encounter} \mathbb{P}_{10} < 1$ regardless of $\alpha$. Physically, this expresses the fact that, for a fixed geometrical cross section and experimental exposure time, we cannot make up enough DM flux to observe a single event with a total SNR large enough to claim a discovery. This cutoff occurs later for larger DM clump radii due to the growing geometrical cross section, as shown in the right panel of \cref{fig:alpha_M}. 
For small values of $M$, the sensitivity to $\alpha$ degrades, and we observe that it scales as approximately $\alpha \propto M^{-3/5}$.  
This scaling arises from a subtle competition between the force, which scales as $F \propto \rho \propto \alpha M$, and the flux, which scales as $N_\mathrm{encounter} \propto M^{-1}$.  

We find that the detector array would be most sensitive to models in which the mass is comparable to the upper threshold from flux considerations. This threshold is greater the larger the radius of the clump, such that higher DM masses are probable. Thus, a detection above the point-like expectation would signal a finite spatial extent for the DM particle. For smaller masses at fixed $\alpha$, the sensitivity degrades as the force scales as $F \propto \alpha M$.  

\section{Summary and Conclusion}
\label{sec:conclusion}

The direct detection of ultraheavy dark matter presents a significant experimental challenge.  
Since the flux of dark matter at the experiment is expected to be small, detectors should be designed to make the most out of every possible encounter.  
Whereas previous studies assumed that ultraheavy dark matter is point-like, it is natural to expect that Planck-mass dark matter would be composite and spatially extended.  
This work represents the first step toward understanding how the finite spatial extent of ultraheavy composite dark matter would impact its prospects for direct detection.  

We focus our study on composite ultraheavy dark matter, having a mass around the Planck scale and a size around a centimeter.  
We consider the direct detection of this dark matter \textit{clump}, assuming that it interacts with a 3-dimensional array of quantum sensors via both gravity and an additional Yukawa force.  
We numerically simulate the response of such a detector during the transit of the dark matter, and in so doing, we infer the range of parameters that would be accessible with such an experimental setup.  
These parameters include the dark matter mass $M$, the dark matter scale radius $R$, the spacing between sensors $d$, the number of sensors in the cubic array $N^3$, the sensor mass $m_s$, the detector temperature $T$, the mechanical damping rate $\gamma$, the quantum squeezing (quantum noise reduction factor) $\xi$, the dimensionless strength parameter of the Yukawa force $\alpha$, and the Yukawa screening length scale $\lambda$.  
We explore different models for the spatial density profile of the dark matter clump: a top-hat, a Gaussian, and an exponential profile.  

Several of our key findings are summarized as follows.
\begin{itemize}
\item 
Detecting composite and ultraheavy dark matter requires a nongravitational interaction: our signal-to-noise ratio never exceeds $1$ if the Yukawa force is absent ($\alpha = 0$); see \cref{fig:alpha_lambda} and \cref{fig:alpha_M}.  
This conclusion is in line with earlier work that studied point-like ultraheavy dark matter \cite{Qin:2025jun}.  

\item
The phenomenology of a sensor array can be split into low $\lambda$ and high $\lambda$ regimes. When the screening length of the Yukawa force is much larger than the scale of the dark matter clump $\lambda \gg R$, we effectively have a long-range interaction that behaves similarly to gravity. A point-like interpretation then generally leads to greater sensitivities. On the other hand, when $\lambda \ll R$, the expected signal for ultraheavy dark matter with a finite spatial extent scales as $\lambda^2$ instead of exponentially---as is the case for a point-like particle. For these small values of $\lambda$, this therefore leads greater sensitivities for finite sized dark matter; see \cref{fig:lambda_R,fig:alpha_lambda}.

\item
If the size $R$ of the dark matter clump is much larger than the scale of the sensor array $L$, the signal is suppressed; see \cref{fig:impulse_R} and \cref{fig:lambda_R}.  
In this limit, the force felt by any individual sensor becomes too small to distinguish from noise.  

\item
For intermediate values of $R$ that are comparable to the sensor spacing $d$, the signal is enhanced and the sensitivity is improved. This means that if one had a strong prior on the size of ultraheavy compact dark matter (in the tabletop range), one could optimize the search for its direct detection by designing a sensor array accordingly.  
\end{itemize}

Our work provides a framework for interpreting composite ultraheavy dark matter models using quantum sensor arrays at large.  
Future work can build upon our study to develop analyses for data coming from sensor array experiments. This could ultimately be used to help discriminate between ultraheavy dark matter candidates given a detection.

\acknowledgments
AJL was supported by the National Science Foundation under Grant No.~PHY-2412797.  DA was jointly supported by NSF Award 2046549 and ERC grant GravNet (ERC-2024-SyG 101167211, DOI: 10.3030/101167211), funded by the European Union. JQ was supported by the U.S.~DOE Office of Science, High Energy Physics, QuantISED program (FWP ERKAP63) and the U.S.~DOE Office of Science, Quantum Science Center. CT was supported by NSF Award 2046549. Views and opinions expressed are however those of the author(s) only and do not necessarily reflect those of the European Union or the European Research Council Executive Agency. Neither the European Union nor the granting authority can be held responsible for them. 

\appendix

\section{Analytical Force Expressions}
\label{sec:app_forces}

For the top-hat, Gaussian, and exponential density profiles in \cref{eq:rho_profs}, the radial integral in \cref{eq:Fvec} can be evaluated analytically to yield the following expressions for the force on a point-like test mass $m_s$ at position $\rvec$: 
\begin{subequations}\label{eq:Fvec_analytical}
\begin{align}
    \text{TH:} & \quad 
    \Fvec(\rvec) 
    = \biggl( - \frac{G m_s \rho_{0,\mrm{T}} \tfrac{4}{3} \pi R^3}{R^2} \rhat \biggr) \begin{cases}
    \frac{r}{R} + \frac{3 \alpha \lambda (r - \lambda) (R + \lambda)}{2 r^2 R} \ee^{\frac{r - R}{\lambda}} + \frac{3 \alpha \lambda (r + \lambda) (R + \lambda)}{2 r^2 R} \, \ee^{-\frac{r+R}{\lambda}} & , \quad r < R \\ 
    \frac{R^2}{r^2} + \frac{3 \alpha \lambda (R - \lambda) (r + \lambda)}{2 r^2 R} \, \ee^{-\frac{r-R}{\lambda}} + \frac{3 \alpha \lambda (r + \lambda) (R + \lambda)}{2 r^2 R} \, \ee^{-\frac{r+R}{\lambda}} & , \quad r > R \\ 
    \end{cases} \\ 
    \text{G:} & \quad 
    \Fvec(\rvec) 
    = \biggl( - \frac{G m_s \rho_{0,\mrm{G}} \tfrac{4}{3} \pi R^3}{R^2} \rhat \biggr) \biggl[
    - \frac{3 R}{r} \, \ee^{-\frac{r^2}{2R^2}} + \frac{3 \sqrt{\pi} R^2}{\sqrt{2} r^2} \, \mathrm{erf}\bigl( \tfrac{r}{\sqrt{2} R} \bigr) 
    - \frac{3 \alpha R}{r} \, \ee^{-\frac{r^2}{2 R^2}} 
    \\ & \hspace{2cm} 
    + \frac{3 \sqrt{\pi} \alpha R^2 (r+\lambda)}{2 \sqrt{2} r^2 \lambda} \, \ee^{-\frac{r}{\lambda}} \ee^{\frac{R^2}{2\lambda^2}} \, \mathrm{erfc}\bigl( \tfrac{-r/R + R/\lambda}{\sqrt{2}} \bigr) 
    \nonumber \\ & \hspace{2cm} 
    + \frac{3 \sqrt{\pi} \alpha R^2 (r - \lambda)}{2 \sqrt{2} r^2 \lambda} \, \ee^{\frac{r}{\lambda}} \ee^{\frac{R^2}{2\lambda^2}} \, \mathrm{erfc}\bigl( \tfrac{r/R + R/\lambda}{\sqrt{2}} \bigr) 
    \biggr] \nonumber \\ 
    \text{E:} & \quad 
    \Fvec(\rvec) 
    = \biggl( - \frac{G m_s \rho_{0,\mrm{E}} \tfrac{4}{3} \pi R^3}{R^2} \rhat \biggr) \biggl[
    \frac{6 R^2}{r^2} 
    + \frac{6 \alpha \lambda^3 R^2 (r + \lambda)}{r^2 (R^2 - \lambda^2)^2} \, \ee^{-\frac{r}{\lambda}} 
    - \frac{3 (r^2 + 2 r R + 2 R^2)}{r^2} \, \ee^{-\frac{r}{R}}
    \nonumber \\ & \hspace{2cm} 
    + \frac{3 \alpha \lambda^2 \bigl( r^2 R^2 - r^2 \lambda^2 - 2 r R \lambda^2 - 2 R^2 \lambda^2 \bigr)}{r^2 (R^2 - \lambda^2)^2} \, \ee^{-\frac{r}{R}} 
    \biggr] \;, 
\end{align}
\end{subequations}
where $\mathrm{erf}(x) = (2/\sqrt{\pi}) \int_0^x \dd x^\prime \, \ee^{-(x^\prime)^2}$ is the error function, and $\mathrm{erfc}(x) = 1 - \mathrm{erf}(x)$ is the complementary error function. 

\section{Numerical Simulation Procedure}
\label{sec:num_alg}

In this appendix we provide a detailed step-by-step algorithm of our numerical simulation procedure.  
We have adapted the algorithm used by the authors of \rref{Qin:2025jun} to account for the dark matter clump's finite spatial extent. 

\begin{enumerate}
    \item[1a.]  \textbf{Dark matter model parameters.}  Select:  $\rho(r)$, $M$, $R$, $\alpha$, and $\lambda$.    
    \item[1b.]  \textbf{Detector model parameters.}  Select:  $m_s$, $R_s$, $d$, and $N$.  Calculate $L = (N-1) d$. We only consider $m_s = 100 \, \mathrm{g}$, $R_s = 1.3 \, \mathrm{cm}$, $d = 10 \, \mathrm{cm}$, $N = 20$, and $L = 1.9\,\mrm{m}$.  
    \item[1c.]  \textbf{Noise model parameters.}  Select: $T$, $\gamma$, $\xi$, and $\tau$.  We only consider $T = 15 \, \mathrm{mK} \approx 1.29 \, \mu\mathrm{eV}$, $\gamma = 4 \pi \times 10^{-12} \, \mathrm{Hz} \approx 1.3 \times 10^{-11} \, \mathrm{Hz}$, $\xi = 15 \, \mathrm{dB} = 10^{1.5} \approx 31.6$, and $\tau \approx 0.24 \, \mathrm{s}$.  
    \item[1d.]  \textbf{Interpolate force.}  We calculate the force $\Fvec(\rvec) = F_r(|\rvec|) \, \rhat$ by evaluating \cref{eq:Fvec_analytical}.  We evaluate $F_r(r)$ at $10^4$ logarithmically-spaced radii in the range $r \in [R_s, 7 (R + L/2) ]$.  From this list, extended by $F_r(R_s)$ at $r=0$, we construct a linear interpolation between adjacent radii. 
    \item[2a.]  \textbf{Detector array.}  Select $Z = 5 (R+L/2)$, draw a random $X^2 = \mathrm{U}\bigl[0, \, (R+L/2)^2 \bigr]$, and draw a random unit vector $\nhat = \mathrm{U}[S^2]$ uniformly distributed on the 2-sphere.  Initialize $N^3$ position 3-vectors $\rvec_{i,0}$ for $i \in \{1, 2, \cdots, N^3\}$ such that they form a three-dimensional cubic array with lattice spacing $d$ centered at $\rvec_0 = X \, \xhat + Z \, \zhat$ with orientation $\nhat$. 
    \item[2b.]  \textbf{Detector trajectory.}  Draw a random speed $v$ from a Maxwell-Boltzmann distribution with $v_\mrm{rms} = v_0 / \sqrt{2}$ and $v_0 = 238\,\mrm{km\,s^{-1}}$.  For each sensor in the array, calculate its position 3-vector at time $t \in [0, t_\mrm{max}]$ as $\rvec_{i}(t) = \rvec_{i,0} - v \zhat t$ with $t_\mrm{max} = 2Z/v$.  This represents the trajectory of the array in the rest frame of the dark matter clump.  We only evaluate the trajectory at discrete time steps $t_j \in \{ 0, \Delta t, 2\Delta t, \cdots, N_\mrm{step} \Delta t \}$ where $\Delta t = \mathrm{floor}(t_\mrm{max} / N_\mrm{step})$.  We take $N_\mrm{step} = 3000$, and we have verified that $\Delta t$ is small enough to reliably resolve $t_\mrm{imp}$. 
    \item[2c.]  \textbf{Calculate SNR.}  For each sensor $i$ in the array, calculate $\Fvec_i(t) = F_r(|\rvec_i(t)|) \, \rhat$ using the interpolating function $F_r(r)$; calculate $\mathcal{S}_{i}$ using \cref{eq:sig} and integrating from $t = 0$ and $t = t_\mrm{max}$; calculate $\mathcal{N}$ using \cref{eq:N}; calculate $\mathrm{SNR}_{i}$ using \cref{eq:SNRi}; and calculate $\mathrm{SNR}$ using \cref{eq:SNR}.  
    \item[3.]  \textbf{Sample trajectories.}  Repeat step \#2 for $N_\mrm{track} = 10^3$ tracks.  For each track, record the calculated $\mathrm{SNR}$.  
    \item[4a.]  \textbf{Detection probability.}  From the resulting distribution over $\mathrm{SNR}$, we calculate the detection probability $\mathbb{P}_{10}$ that $\mathrm{SNR} \geq 10$.  We use $10$ as suggested by \rref{Qin:2023urf}, due to the high trial factor when performing a search for track-like signatures in a sensor array. 
    \item[4b.]  \textbf{Sensitivity projection.}  The projected sensitivity of a detector array to a model of dark matter is defined by the inequality $N_\mrm{encounter} \mathbb{P}_{10} \geq 1$ where $N_\mrm{encounter} = \Phi_\mrm{DM} A t_\mrm{exp}$ is the average number of dark matter clumps with flux $\Phi_\mrm{DM}$ that encounter the detector with geometrical area $A$ during the exposure time $t_\mrm{exp}$.  We take $\Phi_\mrm{DM} = n_\mrm{DM} v_0$, $n_\mrm{DM} = \rho_\mrm{DM} / M$, $\rho_\mrm{DM} \approx 0.3\,\mrm{GeV\,cm^{-3}}$, $v_0 = 238\,\mrm{km\,s^{-1}}$, $A = \pi (R + L/2)^2$, and $t_\mrm{exp} = 5 \, \mathrm{yr}$. If $\mathbb{P}_{10} \geq \mathbb{P}_\mrm{encounter}$ then the initial set of model parameters in step \#1 are said to be within reach of the sensitivity of the experiment.  
\end{enumerate}
We repeat this procedure for several different dark matter model parameters (step \#1) to assess the range of models that are testable. 

\bibliographystyle{JHEP}
\bibliography{refs.bib}

\end{document}